\documentclass[journal,twocolumn,twoside]{IEEEtran}
\usepackage{epsfig,amsfonts,subfigure,amsbsy,bm,mathrsfs}
\usepackage{graphicx,cite,amssymb,amsmath,color}
\usepackage{amsmath,amssymb,amscd,amsthm}
\usepackage{mathtools}
\usepackage[nolist]{acronym}
\usepackage{ctable}
\mathtoolsset{showonlyrefs=true}

\newcommand{\otoprule}{\midrule[\heavyrulewidth]}

\let\originalleft\left
\let\originalright\right
\renewcommand{\left}{\mathopen{}\mathclose\bgroup\originalleft}
\renewcommand{\right}{\aftergroup\egroup\originalright}

\newcommand{\G}{\mathcal{G}}

\newcommand{\na}{\bar{n}_{\mathsf a}}
\newcommand{\PB}{P_{\mathsf F}}
\newcommand{\Pb}{P_{\mathsf b}}
\newcommand{\PBcrdsa}{P^\mathsf{IRSA}_{\mathsf F}}
\newcommand{\Pbcrdsa}{P^\mathsf{IRSA}_{\mathsf P}}
\newcommand{\epss}{\epsilon^*}
\newcommand{\gs}{g^*}
\newcommand{\aepsr}{\alpha_{\epsilon,R}}

\newcommand{\agz}{\alpha_{g,0}}

\newcommand{\C}{\mathcal{C}}
\newcommand{\lC}{\lambda_{\C}}
\newcommand{\rC}{\rho_{\C}}
\newcommand{\msQ}{\mathsf{Q}}

\newcounter{example}

\newcounter{exercice}

\makeatother

\newcommand{\user}{\mathsf{u}}
\newcommand{\slot}{\mathsf{s}}
\newcommand{\vn}{\mathsf{v}}
\newcommand{\cn}{\mathsf{c}}

\begin{acronym}
\acro{LDPC}{low-density parity-check}
\acro{VN}{variable node}
\acro{CN}{check node}
\acro{BP}{belief propagation}
\acro{EXIT}{extrinsic information transfer}
\acro{APP}{a posteriori probability}
\acro{BEC}{binary erasure channel}
\acro{MI}{mutual information}
\acro{PER}{packet error rate}
\acro{PLR}{packet loss rate}
\acro{PLP}{packet loss probability}
\acro{FER}{frame error rate}
\acro{FEP}{frame error probability}
\acro{BEP}{bit error probability}
\acro{CRDSA}{contention resolution diversity slotted ALOHA}
\acro{IRSA}{irregular repetition slotted ALOHA}
\acro{CSA}{coded slotted ALOHA}
\acro{SA}{slotted ALOHA}
\acro{SIC}{successive interference cancellation}
\end{acronym}

\begin{document}

\title{Finite Length Analysis of Irregular Repetition\\ Slotted ALOHA in the Waterfall Region}
\author{Alexandre Graell i Amat, \IEEEmembership{Senior Member, IEEE}, and Gianluigi Liva, \IEEEmembership{Senior Member, IEEE}
\thanks{This work was partially funded by the Swedish Research Council, grant \#2016-04253, and the Ericsson Research Foundation.}
\thanks{A. Graell i Amat is with the
   Department of Electrical Engineering, Chalmers University of Technology,
   	41296 Gothenburg, Sweden (e-mail:  alexandre.graell@chalmers.se).}
	\thanks{G. Liva is with the Institute of Communication and Navigation of the Deutsches Zentrum f\"{u}r Luft- und Raumfahrt (DLR), 82234 Wessling,
   Germany (e-mail: gianluigi.liva@dlr.de).}}

\thispagestyle{empty}
 
\maketitle

\begin{abstract}
A finite length analysis is introduced for irregular repetition slotted ALOHA (IRSA) that enables to accurately estimate its performance in the moderate-to-high packet loss probability regime, i.e., in the so-called waterfall region. The analysis is tailored to the collision channel model, which enables mapping the description of the successive interference cancellation process onto the iterative erasure decoding of low-density parity-check codes. The analysis provides accurate estimates of the packet loss probability of IRSA in the waterfall region as demonstrated by Monte Carlo simulations.
\end{abstract}


\section{Introduction}\label{sec:intro}

\IEEEPARstart{D}{uring} the past decade, a number of efficient random access protocols for massive networks of uncoordinated terminals have been introduced \cite{Casini07,Liva11,Stefanovic13,Paolini15,Iva17,Sandgren17,Vem2017,Clazzer17}. Many of the proposed protocols leverage on \ac{SIC} as a means to improve the throughput with respect to classical random access techniques. Among them, \ac{CRDSA} \cite{Casini07}, \ac{IRSA} \cite{Liva11}, and its variants \cite{Stefanovic13,Paolini15,Iva17,Sandgren17,Vem2017,Clazzer17} gained popularity due to their mild demands in terms of signal processing and their capability to attain substantial throughput gains over the widespread \ac{SA} protocol. 

While the benefits of \ac{IRSA}-like protocols is widely acknowledged, its performance analysis has been historically addressed by means of simulative approaches, with few notable exceptions \cite{delRio14,Ivanov14,Sandgren17}. In \cite{delRio14}, an extensive treatment of the finite length performance of slotted random access protocols is presented which includes, in the model used for analysis, physical layer effects such as the performance of the adopted channel code, capture effects, etc. Due to the ambitious target of the analysis, the approach relies on a mixture of analytical and numerical (i.e., simulative) techniques. In this letter, we take a step back with respect to \cite{delRio14} by addressing the simpler collision model, as for \cite{Ivanov14,Iva17,Sandgren17}. The collision model turns to be accurate when the physical layer implementation does not rely on robust error correcting codes, and hence decoding in the presence of interference is hindered (i.e., no capture effect can be exploited). The collision model has the further advantage of allowing a clean analysis of the interference cancellation process, which enables gaining insights into the iterative process behavior.

In \cite{Ivanov14,Iva17,Sandgren17}, tight analytical approximations to the \ac{PLP} of \ac{IRSA}-like protocols in the \emph{error floor} region, i.e., low channel load regime, were derived. In \cite{Blas17}, an exact finite length analysis of frameless ALOHA was given. However, to the best of our knowledge, analytic approximations to predict the performance of \ac{IRSA} in the so-called \emph{waterfall} region, i.e., moderate-to-high load regime, are still missing. In this letter, we address this problem by providing a way to estimate the \ac{PLP} of \ac{IRSA} in this regime, where the protocol is often operated in practice. The proposed approach leverages on the connection between the \ac{SIC} process and the iterative decoding of \ac{LDPC} codes over the \ac{BEC}. In particular, the finite length scaling analysis of \ac{LDPC} codes over the \ac{BEC} from \cite{Amr09} is adapted to analyze  the packet loss probability of \ac{IRSA}. We show that the developed analytical approximations accurately predict the performance of \ac{IRSA} in the waterfall region. Together with the error floor predictions of \cite{Ivanov14}, they can be used to obtain an accurate performance model for \ac{IRSA} over a wide range of channel loads.

\section{System Model}\label{sec:model}

\begin{figure}[!t]
	\centering
	\includegraphics[width=0.5\columnwidth]{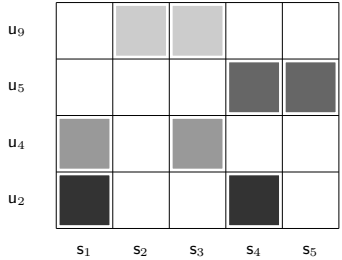}
	\vspace{-2ex}
	\caption{Example of a frame with $m=5$ slots and $4$ active users transmitting $d=2$ copies of the their respective packets.}
	\label{fig:macframe}
	\vspace{-4ex}
\end{figure}

We consider an uncoordinated multiple access scenario with multiple users transmitting to a common receiver based on the \ac{IRSA} protocol, where transmission is organized into frames, each consisting of $m$ slots. We consider a very large (virtually infinite) population of users, of size $n$, which become active sporadically. We denote the set of users as $\{\user_1,\user_2,\ldots,\user_n\}$, and the set of slots as $\{\slot_1,\slot_2,\ldots,\slot_m\}$.  Users are slot- and frame-synchronous and each user attempts at most one packet transmission per frame. A user transmitting in a frame is referred to as active. According to the \ac{IRSA} protocol, each active user transmits a number of copies, $d$, of its packet within a frame according to a distribution
\begin{align}
\Lambda(x)=\sum_d \Lambda_dx^d
\end{align} 
where $\Lambda_d$ is the probability that a user transmits $d$ copies. The $d$ copies are transmitted in $d$ distinct slots chosen uniformly at random. We denote by $\na$ the expected number of users that are active in a given frame. The expected channel load is then
\begin{align}
g=\na/m.
\end{align}
An example of a frame with $n=10$ and $m=5$ is depicted in Fig.~\ref{fig:macframe}. Out of the $10$ users, $4$ users ($\user_2,\user_4,\user_5$, and $\user_9$) are active and transmit two copies of their packets in randomly-selected slots. Each packet is equipped with a pointer to the position of its copies. We restrict to the collision channel model. The receiver stores an observation of the full frame and decodes the packets by iterating the following steps: i) For each slot containing a single packet, the packet is decoded; ii) For each decoded packet, the pointer to its copies is extracted, and the interference contribution caused by the packet copies is removed from the frame.
The \ac{SIC} process is iterated until no further packets can be decoded. The model, despite being simple, can be used to obtain a first estimate of the performance of \ac{IRSA} under more realistic assumptions \cite{Casini07,Liva11}.

\section{Connection Between \ac{IRSA} and High-Rate LDPC Codes}

In \cite{Pao17}, a link between the \ac{IRSA} scheme with a large population of users and high-rate LDPC codes for transmission over the \ac{BEC} was highlighted. Here, we exploit the link to estimate the performance of \ac{IRSA} in the waterfall region, borrowing tools for the finite length analysis of \ac{LDPC} codes.

\begin{figure}[!t]
	\centering
	\includegraphics[width=0.75\columnwidth]{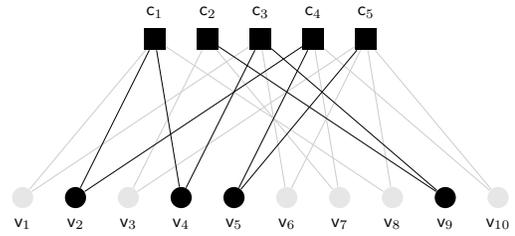}
	\vspace{-1.5ex}
	\caption{Bipartite graph representation of the frame depicted in Fig.~\ref{fig:macframe}.}
	\label{fig:example}
	\vspace{-3ex}
\end{figure}

Consider the bipartite graph representation of \ac{IRSA}, where each of the $n$ users is represented by a \ac{VN} and each of the $m$ slots of a frame is represented by a \ac{CN}. Let us denote by $\{\vn_1,\vn_2,\ldots,\vn_n\}$ the set of $n$ \acp{VN} and by $\{\cn_1,\cn_2,\ldots,\cn_m\}$ the set of $m$ \acp{CN}. We have that $\vn_j$ is connected to $\cn_i$ if and only if user $\user_j$ selects slot $\slot_i$ for the transmission of its packet (copy). The bipartite graph corresponding to the frame of Fig.~\ref{fig:macframe} is provided in Fig.~\ref{fig:example}. In Fig.~\ref{fig:example}, the \acp{VN} associated to active users are depicted as dark circles, whereas the \acp{VN} associated to inactive users are shown in light gray. Observe that edges are also drawn between \acp{CN} and \acp{VN} associated to inactive users. Obviously, inactive users do not cause any interference in the frame. Nevertheless, the inclusion of edges connected to their associated \acp{VN} turns to be instrumental to the following observation: The access scheme described in Section~\ref{sec:model} can be cast in an equivalent manner by assuming that each of the $n$ users first picks a repetition degree $d$ according to $\Lambda(x)$, and then it selects $d$ slots at random on which the user \emph{may} transmit $d$ copies of a packet. If a packet is available at the user, then the user proceeds with the transmission (becoming active), otherwise the user remains silent during the frame. Hence, edges connected to a \ac{VN} associated to an inactive user refer to the slot selection performed by the inactive user. Evidently, the behavior of the \ac{SIC} algorithm described in Section~\ref{sec:model} is not affected in any way by the slot selection performed by inactive users.
Since each users transmits $d$ copies of its packet according to $\Lambda(x)$, the resulting bipartite graph has edge-perspective \ac{VN} degree distribution
\begin{align}
\lambda(x)=\sum_d\lambda_dx^{d-1} \label{eq:VNdd}
\end{align}
where $\lambda_d=\Lambda_dd/\sum_d\Lambda_d d$.
For large $m$,  the number of transmissions in a slot follows a Poisson distribution, i.e., the edge-perspective \ac{CN} degree distribution is
\begin{align}
\rho(x)=\exp(-g\Lambda'(1)(1-x)) \label{eq:CNdd}
\end{align}
where $\Lambda'(x)$ denotes the derivative of $\Lambda(x)$.

It is interesting to observe that the resulting bipartite graph corresponds to that of a high-rate Poisson \ac{LDPC} code ensemble $\mathcal{C}$ with code length $n$, nominal rate $R$, VN degree distribution $\lC(x)=\lambda(x)$, and Poisson CN degree distribution
\begin{align}
\rC(x)=\exp\left(\frac{-\Lambda'(1)(1-x)}{R}\right)
\end{align}
where
\vspace{-0.1cm}
\begin{align}\label{eq:rate}
R=(n-m)/n.
\end{align}
We denote such a Poisson LDPC code ensemble by $\mathcal{C}(n,\lC(x),R)$. Note that if for a fixed $m$ the population size grows very large, i.e., $n\rightarrow\infty$, the code rate of the \emph{equivalent} Poisson LDPC code ensemble tends to one, $R\rightarrow 1$.

Consider a graph $\G$ with $n$ \acp{VN} and $m$ \acp{CN} drawn according to \eqref{eq:VNdd}, \eqref{eq:CNdd}. Assume the graph to represent an instance of transmission with \ac{IRSA}, and denote by $\mathcal{S}_{\mathsf a}$ the set of the indices of the active users. The iterative recovery process of the users with indices in $\mathcal{S}_{\mathsf a}$ is equivalent to the erasure decoding of an \ac{LDPC} code with the same bipartite graph $\G$, where the \acp{VN} associated to erased codeword bits have indices in $\mathcal{S}_{\mathsf a}$. Hence, the analysis of the iterative \ac{SIC} process for \ac{IRSA} with edge-perspective degree distribution $\lambda(x)$ over a frame with $m$ slots and $n$ users can be cast as the analysis of the iterative erasure decoding for an \ac{LDPC} code picked from $\mathcal{C}(n,\lC(x),R)$ (with $R$ given by \eqref{eq:rate}) over the \ac{BEC} with erasure probability $\epsilon$.
The channel load of the \ac{IRSA} system can be written as a function of the erasure probability of the \ac{BEC} as
\begin{align}\label{eq:g}
g=\frac{\na}{m}=\frac{\epsilon n}{m}=\frac{\epsilon}{1-R}.
\end{align}
In the asymptotic regime of infinitely large number of slots, $m\rightarrow\infty$, the iterative decoding performance of \ac{IRSA} shows a threshold behavior. We denote by $g^*$ the \ac{BP} decoding threshold of \ac{IRSA}, i.e., $g^*$ is the maximum channel load for which the probability of not resolving a user is vanishing small in the limit of infinitely large population size and frame length (with constant ratio). Let $\epss_R$ be the \ac{BP} threshold of the equivalent LDPC code ensemble $\mathcal{C}(n,\lC(x),R)$. Following \eqref{eq:g}, the decoding threshold of \ac{IRSA} can then be expressed in terms of $\epss_R$ as
\begin{align}\label{eq:gstar}
\gs=\frac{\epss_R n}{m}=\frac{\epss_R}{1-R}.
\end{align}

\section{Finite Length Scaling}\label{sec:scaling}

Exploiting the analogy between \ac{IRSA} and \ac{LDPC} code ensembles discussed in the previous section, in the following we provide approximations to the \ac{FEP} and \ac{PLP} of \ac{IRSA} in the waterfall region. In particular, we adapt the finite-length scaling framework of \cite{Amr09} to \ac{IRSA}.

\subsection{Frame Error Probability}

The waterfall region performance of an \ac{LDPC} code ensemble $\mathcal{C}(n,\lC(x),R)$ over the \ac{BEC} can be characterized using the finite-length scaling framework of \cite{Amr09}. In particular, the \ac{FEP}, denoted by $\PB^{\C}$, can be expressed as a function of $n$ and the channel erasure probability $\epsilon$ as \cite[eq.~(7)]{Amr09}
\begin{align}\label{eq:PB}
\PB^{\C}\approx \msQ\left(\frac{\sqrt{n}\left(\epss_R-\beta_R n^{-2/3}-\epsilon\right)}{\aepsr}\right)
\end{align}
where $\aepsr=\sqrt{{\alpha^2_R}+\epsilon(1-\epsilon)}$ is a parameter that depends on the code rate $R$ and the channel erasure probability $\epsilon$, whereas $\beta_R$ and $\alpha_R$ are constants that depend only on the code rate (i.e., they are independent of $\epsilon$). In \eqref{eq:PB} $\msQ(x)$ is the tail probability of the standard normal distribution. Furthermore, $\epss_R$, $\aepsr$, and $\beta_R$ can be expressed as \cite{Amr09}
\begin{align}
\epss_R&=(1-R)\epss_0 \label{eq:epssr}\\
\alpha_R&=(1-R)^{1/2}\alpha_0 \label{eq:alphar}\\
\beta_R&=(1-R)^{1/3}\beta_0\label{eq:betar}
\end{align}
where $\epss_0$ is the \ac{BP} threshold and $\alpha_0$ and $\beta_0$ scaling parameters for the zero-rate ensemble $\mathcal{C}(n,\lC(x),R=0)$. The values of $\epss_0$, $\alpha_0$, and $\beta_0$ can be found in \cite[Table~II]{Amr09} for regular VN degree distribution $\lC(x)=x^{d-1}$ and several values of $d$. 

Using the analogy between \ac{IRSA} and LDPC code ensembles, we can use \eqref{eq:PB} with some modifications to predict the finite length performance of \ac{IRSA} in the waterfall region. First, note that by using \eqref{eq:epssr} in \eqref{eq:gstar} it follows that $\gs=\epss_0$, i.e., the \ac{BP} threshold of the \ac{IRSA} scheme  is equal to the \ac{BP} threshold of the ensemble $\mathcal{C}(n,\lC(x),0)$ with $\lC(x)=\lambda(x)$. Now, using $n={m}/(1-R)$ (from \eqref{eq:rate}) together with \eqref{eq:epssr}, \eqref{eq:alphar}, and \eqref{eq:betar} in \eqref{eq:PB}, the \ac{FEP} of \ac{IRSA} can be written in terms of $m$, $\gs$, and $g$ as
\begin{align}\label{eq:PfCRDSA}
\PBcrdsa\approx \msQ\left(\frac{\sqrt{m}\left(\gs-\beta_0m^{-2/3}-g\right)}{\agz}\right)
\end{align}
where $\agz=\sqrt{\alpha^2_0+g(1-(1-R)g)}$. Letting $n\rightarrow\infty$, we have $R\rightarrow 1$ and it follows that $\agz=\sqrt{\alpha^2_0+g}$, yielding
\begin{align}\label{eq:PfCRDSAb}
\PBcrdsa\approx \msQ\left(\frac{\sqrt{m}\left(\gs-\beta_0m^{-2/3}-g\right)}{\sqrt{\alpha^2_0+g}}\right).
\end{align}
The value of $g^*$ and the scaling parameters $\alpha_0$ and $\beta_0$ for $\Lambda(x)=x^d$ (i.e., \ac{CRDSA}) and $d=3,4,$ and $5$, and for $\Lambda_1(x)=0.5x^4+0.5x^8$ and $\Lambda_2(x)=0.86x^3+0.14x^8$ are given in Table~\ref{tab:ScalingParameters}. For the irregular distributions, the scaling parameters have been computed using the method in \cite{Amr06b}.
\begin{table}[t]
	\addtolength{\tabcolsep}{-0.7mm}
	\scriptsize
	\caption{Scaling parameters for \ac{IRSA}}
	\vspace{-4.5ex}
	\begin{center}\begin{tabular}{cccccc}
			\hline
			\toprule
			$\Lambda(x)$  &  $g^*$  & $\alpha_0$ & $\beta_0$ & $\gamma$ \\
			\otoprule
	     	$x^3$  &  $0.818469$  & $0.497867$ & $0.964528$ & $0.783499$ \\[0.5mm]
		$x^4$  &  $0.772280$  & $0.409321$ & $0.827849$ & $0.906054$\\[0.5mm]
		$x^5$  &  $0.701780$  & $0.375892$ & $0.760593$ & $0.961253$ \\[0.5mm]
		$\Lambda_1(x)$  &  $0.661090$  & $0.404986$ & $0.849037$ &  $0.982040$ \\[0.5mm]
		$\Lambda_2(x)$  & $0.851325$  & $0.496301$ & $1.50477$ & $0.835418$  \\[0.5mm]
		\bottomrule
		\end{tabular} \end{center}
		\label{tab:ScalingParameters} 
		\vspace{-7ex}
	\end{table}

Note that \eqref{eq:PfCRDSAb} is capable of modeling the performance of \ac{IRSA} down to a moderate \ac{FEP} (i.e., in the so-called waterfall region of the \ac{FEP} curve, at moderate-to-high channel loads). At low  \ac{FEP} (i.e., low channel load), the performance of \ac{IRSA} exhibits a typical error floor phenomenon which may be predicted through combinatorial analysis methods as shown in \cite{Ivanov14,Iva17,Sandgren17}. We will see in Section \ref{sec:NumericalResults} how the combination of the two approaches can provide a tight estimate of the \ac{FEP} over the whole range of channel loads.

\subsection{Packet Loss Probability}

In \cite{Amr09}, the \ac{BEP} of \ac{LDPC} codes was approximated as
\begin{align}
\label{eq:Pbb}
\Pb^{\C}\approx \nu^* \PB^{\C}
\end{align}
where $\nu^*$ is the fraction of \acp{VN} that cannot be decoded at the BP threshold, in the limit of an asymptotically large block length $n$.  Thus, $\nu^*$ is a constant that does not depend on $\epsilon$, and $\Pb^{\C}$ is simply obtained by scaling $\PB^{\C}$. Our numerical results suggest that the heuristic approximation in \eqref{eq:Pbb} is accurate for relatively large block lengths but does not predict well the performance for very short block lengths. Since \ac{IRSA} is typically operated with frames composed by a few tens (or hundreds) of slots, the scaling law \eqref{eq:Pbb} shall be suitably modified.  Indeed, there is no evidence that $\eqref{eq:Pbb}$ accurately predicts well the expected \ac{BEP} of \ac{LDPC} code ensembles for short block lengths.\footnote{The only result on the \ac{BEP} in \cite{Amr09}  is unfortunately not correct, in the sense that the reported figure is in fact a reproduction of the \ac{FEP} result.} Here, we propose the following approximation of the \ac{PLP} of \ac{IRSA},
\begin{align}
\Pbcrdsa\approx \gamma \PBcrdsa
 \approx \gamma \msQ\left(\frac{\sqrt{m}\left(\gs-\beta_0m^{-2/3}-g\right)}{\agz}\right)
\label{eq:PbCRDSA}
\end{align}
where $\gamma$ is the   \ac{PLP} for $g\rightarrow 1$ computed via density evolution \cite{Liva11}. The value of $\gamma$ for several distributions $\Lambda(x)$ is given in Table~\ref{tab:ScalingParameters}.  

\section{Numerical Results}
\label{sec:NumericalResults}

\begin{figure}[!t]
	\centering
	\includegraphics[width=0.9\columnwidth]{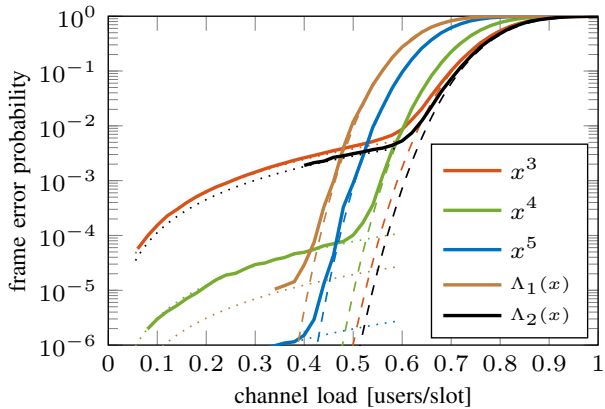}
	\vspace{-2ex}
	\caption{Finite length approximation \eqref{eq:PfCRDSAb} (dashed lines), simulation results (solid lines), and error floor approximation (dotted lines) of the frame error probability of \ac{IRSA} for $m=200$.}
	\label{fig:FER_n200}
	\vspace{-2.5ex}
\end{figure}

In Fig.~\ref{fig:FER_n200}, we plot the \ac{FEP} estimate for \ac{IRSA} according to \eqref{eq:PfCRDSAb} and the parameters in Table~\ref{tab:ScalingParameters} (dashed curves) as a function of the channel load $g$, together with simulation results for the \ac{FER} (solid curves) for $m=200$ and the distributions $\Lambda(x)=x^{d}$ with $d=3,4$, and $5$, $\Lambda_1(x)$, and $\Lambda_2(x)$. As can be observed, the \ac{FEP} estimates predict very accurately the performance of \ac{IRSA} in the waterfall region. At a certain channel load, the simulated curves diverge from the analytical curves. This corresponds to the region of channel load values where the slope of the \ac{FER} changes and the \ac{FER} approaches the error floor. In the figure, we plot the approximation to the error floor derived in \cite{Ivanov14,Iva17,Sandgren17}.\footnote{In \cite{Ivanov14,Iva17,Sandgren17}, analytical expressions to accurately predict the error floor of the \ac{PLP} were derived. The analysis can be extended to the \ac{FEP} in a straightforward manner.}
\begin{figure}[!t]
	\centering
	\includegraphics[width=0.9\columnwidth]{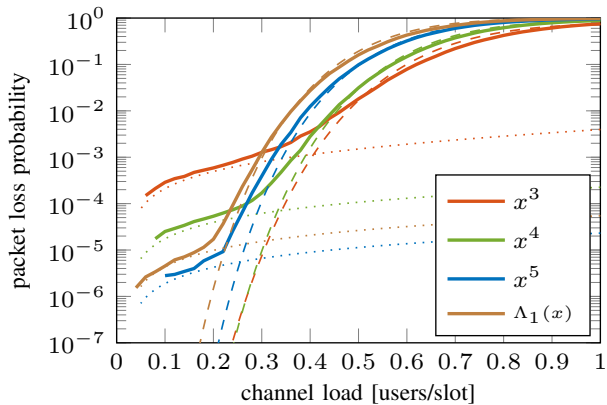}
	\vspace{-2ex}
	\caption{Finite length approximation \eqref{eq:PbCRDSA} (dashed lines), simulation results (solid lines), and error floor approximation (dotted lines) of the packet loss probability \cite{Ivanov14,Iva17} of \ac{IRSA} for $m=50$.} 
	\label{fig:FL_PLR_n50}
	\vspace{-3ex}
\end{figure}

\begin{figure}[!t]
	\centering
	\includegraphics[width=0.9\columnwidth]{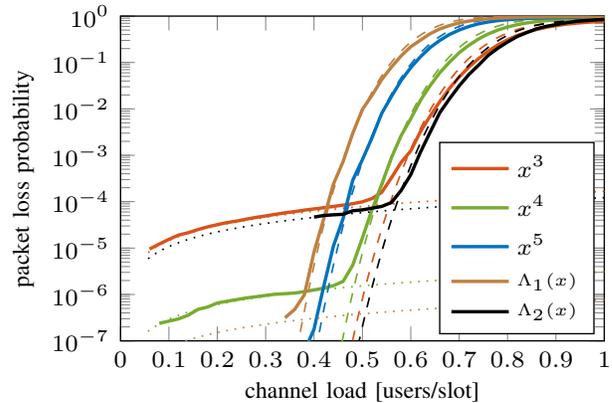}
	\vspace{-2ex}
	\caption{Finite length approximation \eqref{eq:PbCRDSA} (dashed lines), simulation results (solid lines), and error floor approximation (dotted lines) of the packet loss probability \cite{Ivanov14,Iva17} of \ac{IRSA} for $m=200$.}
	\label{fig:FL_PLR_n200}
	\vspace{-3ex}
\end{figure}

In Figs.~\ref{fig:FL_PLR_n50} and~\ref{fig:FL_PLR_n200}, we plot the \ac{PLP} estimate for \ac{IRSA} obtained using \eqref{eq:PbCRDSA} and the parameters in Table~\ref{tab:ScalingParameters}  (dashed curves)
as a function of the channel load $g$
together with simulation results (solid curves) for $m=50$ and $m=200$, respectively. The analytical curves slightly  overestimate the \ac{PLP} in the region where the curve bends down to the waterfall. However, there is a significant agreement with the simulation results, even for small $m$.
On the same charts, the approximation to the error floor performance derived in \cite{Ivanov14,Iva17,Sandgren17} is provided again.

\section{Conclusion}\label{sec:conclusions}

We derived analytical approximations to the packet loss probability of \ac{IRSA} in the medium-to-high load regime. The derived approximations give tight predictions of the performance of \ac{IRSA} in this region. Together with the approximations for the error floor previously derived, they allow for an accurate characterization of the performance of \ac{IRSA} over a wide range of channel loads and can be used to optimize the degree distribution for a given frame length and PLR. The proposed analysis is also applicable to \ac{CRDSA}, which can be seen as an instance of \ac{IRSA} with regular VN degree.

\vspace{-0.1cm}
\section*{Acknowledgment}
The authors are grateful to F.~Clazzer for providing the simulation results for \ac{CRDSA}.

\vspace{-0.1cm}

\bibliographystyle{IEEEtran}

\end{document}